\documentclass[english,journal=jctcce,manuscript=article,articletitle=true]{achemso}
\usepackage[T1]{fontenc}
\usepackage[latin9]{inputenc}
\usepackage{array}
\usepackage{multirow}
\usepackage{amsmath}
\usepackage{amssymb}
\usepackage{graphicx}

\makeatletter

\title{Stochastic $\text{GW}$ calculations for molecules}
\author{Vojt\v{e}ch Vl\v{c}ek}
\email{vojtech@chem.ucla.edu}
\affiliation{Department of Chemistry and Biochemistry, University of California,
Los Angeles California 90095, U.S.A.}
\alsoaffiliation{Fritz Haber Center for Molecular Dynamics, Institute of Chemistry,
The Hebrew University of Jerusalem, Jerusalem 91904, Israel}
\author{Eran Rabani}
\email{eran.rabani@berkeley.edu}
\affiliation{Department of Chemistry, University of California and Materials Science
Division, Lawrence Berkeley National Laboratory, Berkeley, California
94720, USA}
\alsoaffiliation{The Raymond and Beverly Sackler Center for Computational Molecular
and Materials Science, Tel Aviv University, Tel Aviv, Israel 69978}
\author{Daniel Neuhauser}
\email{dxn@chem.ucla.edu}
\affiliation{Department of Chemistry and Biochemistry, University of California,
Los Angeles California 90095, U.S.A.}
\author{Roi Baer}
\email{roi.baer@huji.ac.il}
\affiliation{Fritz Haber Center for Molecular Dynamics, Institute of Chemistry,
The Hebrew University of Jerusalem, Jerusalem 91904, Israel}
\providecommand{\tabularnewline}{\\}

\SectionNumbersOn

\makeatother

\usepackage{babel}
\begin{document}
\begin{abstract}
Quasiparticle (QP) excitations are extremely important for understanding
and predicting charge transfer and transport in molecules, nanostructures
and extended systems. Since density functional theory (DFT) within
the Kohn-Sham (KS) formulation does not provide reliable QP energies,
many-body perturbation techniques such as the GW approximation are
essential. The main practical drawback of GW implementations is the
high computational scaling with system size, prohibiting its use in
extended, open boundary systems with many dozens of electrons or more.
Recently, a stochastic formulation of GW (sGW) was presented {[}\emph{Phys.
Rev. Lett.} \textbf{113, }076402 (2014){]} with a near-linear-scaling
complexity, illustrated for a series of silicon nanocrystals reaching
systems of more than $3000$ electrons. This advance provides a route
for many-body calculations on very larges systems that were impossible
with previous approaches. While earlier we have shown the gentle scaling
of sGW, its accuracy was not extensively demonstrated. Therefore,
we show that this new sGW approach is very accurate by calculating
the ionization energies of a group of sufficiently small molecules
where a comparison to other GW codes is still possible. Using a set
of 10 such molecules, we demonstrate that sGW provides reliable vertical
ionization energies in close agreement with benchmark deterministic
GW results {[}J. Chem. Theory Comput, \textbf{11}, 5665 (2015){]},
with mean (absolute) deviation of 0.05 and 0.09eV. For completeness,
we also provide a detailed review of the sGW theory and numerical
implementation. 
\end{abstract}

\section{\label{sec:intro}Introduction}

First-principles electronic structure calculations play a central
role in predicting and understanding the behavior of molecules, nanostructures
and materials. For the ground state, the methods of choice are density
functional theory,\cite{Hohenberg1964,Kohn1965} Hartree\textendash Fock
(HF), and to some extent post HF techniques such as the M\"{o}ller\textendash Plesset
perturbation theory. Ground state calculations are routinely possible
for extended, finite systems due to fast numerical electronic structure
solvers and the increases in computational power (see Ref.~\citenum{VandeVondeleBorstnikHutter2012}
and references therein).

For charge (quasiparticle) and neutral (optical) excitations, the
situation is more complex, and most if not all methods are limited
to either small molecules or to periodic crystals with a relatively
small unit cell.\cite{Hybertsen1985,Steinbeck1999,Shishkin2007,Rostgaard2010,Foerster2011,Faber2011,Blase2011a,Deslippe2012,Marom2012,Caruso2012,vanSetten2013,Pham2013,Govoni2015,Kaplan2016}
While DFT is a theory for the ground state, recent developments using
hybrid functionals\cite{Heyd2003,Baer2010a,Kronik2012} extend the
use of DFT to describe QP excitations, even in system with thousands
of electrons.\cite{Neuhauser2015} However, the description of the
QP excitations within DFT hybrids lacks dynamical effects, such as
screening and lifetime of the QPs. An alternative for describing electronic
excitations is the many-body perturbation theory within the GW approximation
for charged QPs~\cite{Hedin1965,Hybertsen1985,Hybertsen1986,Aryasetiawan1998,Onida2002,Friedrich2006}
and BSE for QPs associated with neutral excitations.\cite{Rohlfing2000,Onida2002,Benedict2003,Tiago2006}
Both approaches scale steeply with system size and therefore are very
expensive for large systems. 

Recently, we developed a stochastic approach for both flavors, stochastic
GW ($\text{sGW}$)~\cite{Neuhauser2014} and the stochastic Bethe-Salpeter
equation ($\text{sBSE}$) approach.\cite{Rabani2015} The former scales
near-linearly and the latter scales quadratically with system size.
Both stochastic methods extend significantly the size of systems that
can be studied within many-body perturbation techniques. Furthermore,
of the two, sGW is fully \emph{ab initio} and can be therefore compared
to other GW formulations.

In this paper we assess the accuracy and convergence of sGW versus
other well-established codes. This is important since the GW literature
contains a wide spread of results for the same systems.\cite{vanSetten2015}
While the theoretical foundations of sGW are solid,\cite{Neuhauser2014}
the approach has not been tested extensively for systems that are
small enough so they can be studied by conventional deterministic
programs. For this comparison, we selected a group of $10$ small
molecules containing first row atoms (for which experimental geometries
and vertical ionization potentials are available) and compared the
sGW results for vertical ionization energies to those of well tested
\cite{vanSetten2015} state-of-the-art deterministic methods based
on the GW implementation within TURBOMOLE \cite{furche2014turbomole,vanSetten2013}
and FHI-aims.\cite{Blum2009,Ren2012a}

In Section~\ref{sec:Stochastic-formulation-of} we review the $\text{sGW}$
formalism.\cite{Neuhauser2014} In Section~\ref{sec:Results} we
summarize the results for the subset of $10$ molecules. Summary and
conclusions follow in Section~\ref{sec:Summary}.

\section{\label{sec:Stochastic-formulation-of}Stochastic formulation of the
$\text{\ensuremath{G_{0}W_{0}}}$ approximation}

\subsection{$\boldsymbol{\text{\ensuremath{G_{0}W_{0}}}}$ in the energy domain}

It is possible to write a formal equation for the QP Dyson orbitals
$\psi_{n}^{QP}\left(\boldsymbol{r}\right)$ and energies $\varepsilon_{n}^{QP}$:
\begin{equation}
-\frac{\hbar^{2}}{2m_{e}}\nabla^{2}\psi_{n}^{QP}\left(\boldsymbol{r}\right)+v_{\text{ext}}\left(\boldsymbol{r}\right)\psi_{n}^{{\rm QP}}\left(\boldsymbol{r}\right)+v_{\text{H}}\left(\boldsymbol{r}\right)\psi_{n}^{QP}\left(\boldsymbol{r}\right)+\int\tilde{\Sigma}\left(\boldsymbol{r},\boldsymbol{r}^{\prime},\varepsilon_{n}^{QP}\right)\psi_{n}^{QP}\left(\boldsymbol{r}^{\prime}\right){\rm d}\boldsymbol{r}^{\prime}=\varepsilon_{n}^{QP}\psi_{n}^{QP}\left(\boldsymbol{r}\right)\label{eq:QP}
\end{equation}
which is similar to a Schr�dinger equation, containing kinetic energy
and external potential energy ($v_{\text{ext}}\left(\boldsymbol{r}\right)$)
operators as well as a mean electrostatic or Hartree potential 
\begin{align}
v_{\text{H}}\left(\boldsymbol{r}\right) & =\int n\left(\boldsymbol{r}^{\prime}\right)u_{C}\left(\left|\boldsymbol{r}-\boldsymbol{r}^{\prime}\right|\right)d\boldsymbol{r}^{\prime},\label{eq:V-Hartee}
\end{align}
where $n\left(\mathbf{r}\right)$ is the ground-state density of the
$N$-electron system and $u_{C}\left(r\right)=\frac{e^{2}}{4\pi\epsilon_{0}r}$
is the bare Coulomb potential energy. This equation also contains
a non-local energy-dependent self-energy term $\tilde{\Sigma}\left(\boldsymbol{r},\boldsymbol{r}^{\prime},\omega\right)$
which incorporates the many-body exchange and correlation effects
into the system. Eq.~(\ref{eq:QP}) is exact, but requires the knowledge
of the self-energy which cannot be obtained without imposing approximations.
One commonly used approach is based on the GW approximation.\cite{Hedin1965}
However, even this theory is extremely expensive computationally and
a further simplification is required leading to the so-called $G_{0}W_{0}$
approximation

\begin{eqnarray}
\tilde{\Sigma}\left(\boldsymbol{r},\boldsymbol{r}',\omega\right) & = & i\int_{-\infty}^{\infty}\frac{d\omega^{\prime}}{2\pi}\tilde{G}_{0}\left(\boldsymbol{r},\boldsymbol{r}',\omega+\omega^{\prime}\right)\tilde{W}_{0}\left(\boldsymbol{r},\boldsymbol{r}',\omega^{\prime}\right).\label{eq:G0W0(w)}
\end{eqnarray}
$\tilde{G}_{0}\left(\mathbf{r},\mathbf{r}^{\prime},\omega\right)$
is a time-ordered Green's function given by: 
\begin{equation}
\tilde{G}_{0}\left(\boldsymbol{r},\boldsymbol{r}',\omega\right)=\lim_{\eta\to0^{+}}\hbar\sum_{n}\phi_{n}^{KS}\left({\bf r}\right)\phi_{n}^{KS}\left({\bf r^{\prime}}\right)\left[\frac{f_{n}}{\hbar\omega-\varepsilon_{n}^{KS}-i\eta}+\frac{1-f_{n}}{\hbar\omega-\varepsilon_{n}^{KS}+i\eta}\right],\label{eq:G0(w)}
\end{equation}
within a Kohn-Sham (KS) DFT starting point.\cite{Hohenberg1964,Kohn1965}
$\phi_{n}^{KS}\left({\bf r}\right)$ and $\varepsilon_{n}^{KS}$ are
the real KS eigenstates and eigenvalues, respectively, of the KS Hamiltonian
(henceforth, we use atomic units where\textbf{ $\hbar=m_{e}=e=4\pi\epsilon_{0}=1$})
\begin{equation}
\hat{h}_{KS}=-\frac{1}{2}\nabla^{2}+v_{{\rm ext}}\left(\boldsymbol{r}\right)+v_{{\rm H}}\left(\boldsymbol{r}\right)+v_{\text{xc}}\left(\boldsymbol{r}\right),
\end{equation}
and $v_{\text{xc}}\left(\boldsymbol{r}\right)$ is the exchange-correlation
potential that depends on the ground state density, $n\left(\boldsymbol{r}\right)$.
In Eq.~\ref{eq:G0(w)}, $f_{n}$ is the occupation of the KS level
$n$. In Eq.~(\ref{eq:G0W0(w)}), $\tilde{W}_{0}\left(\boldsymbol{r},\boldsymbol{r}',\omega^{\prime}\right)$
is the \emph{time-ordered }screened Coulomb potential defined as
\begin{equation}
\tilde{W}_{0}\left(\boldsymbol{r},\boldsymbol{r}',\omega\right)=\int\epsilon^{-1}\left(\boldsymbol{r},\boldsymbol{r}'',\omega\right)u_{C}\left(\left|\boldsymbol{r}''-\boldsymbol{r}'\right|\right)d\boldsymbol{r}'',\label{eq:W0}
\end{equation}
where $\epsilon^{-1}\left(\boldsymbol{r},\boldsymbol{r}',\omega\right)=\delta\left(\boldsymbol{r}-\boldsymbol{r}'\right)+\int u_{C}\left(\left|\boldsymbol{r}-\boldsymbol{r}''\right|\right)\tilde{\chi}\left(\boldsymbol{r}'',\boldsymbol{r}',\omega\right)d\boldsymbol{r}''$
is the frequency dependent inverse dielectric function and $\tilde{\chi}\left(\boldsymbol{r},\boldsymbol{r}',\omega\right)$
is the reducible polarizability. 

Once the self-energy is generated via Eqs.~(\ref{eq:G0W0(w)})-(\ref{eq:W0})
the QP energies of Eq.~(\ref{eq:QP}) can be estimated perturbatively,
as a correction to the KS orbital energies. To first order:\cite{Hybertsen1985,Hybertsen1986}

\begin{align}
\varepsilon_{n}^{QP} & =\varepsilon_{n}^{KS}-V_{XC}+\tilde{\Sigma}_{n}\left(\varepsilon_{n}^{QP}\right),\label{eq:perturbativeQPE}
\end{align}
where $V_{XC}=\int v_{XC}\left(\boldsymbol{r}\right)\left|\phi_{n}^{KS}\left(\boldsymbol{r}\right)\right|^{2}d\boldsymbol{r}$
is the expectation value of the exchange-correlation potential, and
$\tilde{\Sigma}_{n}\left(\omega\right)$ is the self energy expectation
value at a frequency $\omega$:
\begin{equation}
\tilde{\Sigma}_{n}\left(\omega\right)=\iint\phi_{n}^{KS}\left(\boldsymbol{r}\right)\tilde{\Sigma}\left(\boldsymbol{r},\boldsymbol{r}',\omega\right)\phi_{n}^{KS}\left(\boldsymbol{r}'\right)d\boldsymbol{r}d\boldsymbol{r}'.\label{eq:Sigma-ii(w)}
\end{equation}

\subsection{$\boldsymbol{\text{\ensuremath{G_{0}W_{0}}}}$ in the time domain}

The computational challenge of $\text{\ensuremath{G_{0}}}\text{\ensuremath{W_{0}}}$
is to estimate the frequency-dependent function $\tilde{\Sigma}_{n}\left(\omega\right)$
involving integration over $6$-dimensional quantities. A simplification
is achieved when we Fourier transform to the time domain
\begin{equation}
\Sigma_{n}\left(t\right)\equiv\int_{-\infty}^{\infty}\tilde{\Sigma}_{n}\left(\omega\right)e^{-i\omega t}\frac{d\omega}{2\pi},
\end{equation}
since the self-energy in the time domain is a simple \emph{product
}of the time domain Green's function and screened potential 
\begin{align}
\Sigma\left(\boldsymbol{r},\boldsymbol{r}',t\right) & =iG_{0}\left(\boldsymbol{r},\boldsymbol{r}',t\right)W_{0}\left(\boldsymbol{r},\boldsymbol{r}',t^{+}\right),\label{eq:G0W0(t)}
\end{align}
instead of the convolution in Eq.~(\ref{eq:G0W0(w)}). In Eq.~(\ref{eq:G0W0(t)}),
$t^{+}$ is a time infinitesimally later than $t$ and $G_{0}\left(\boldsymbol{r},\boldsymbol{r}',t\right)$
is the Fourier transform of $\tilde{G}_{0}\left(\boldsymbol{r},\boldsymbol{r}',\omega\right)$,
given by: 
\begin{equation}
iG_{0}\left(\boldsymbol{r},\boldsymbol{r}',t\right)=\sum_{n}\phi_{n}^{KS}\left(\boldsymbol{r}\right)\phi_{n}^{KS}\left(\boldsymbol{r}'\right)e^{-i\varepsilon_{n}^{KS}t/\hbar}\left[\left(1-f_{n}\right)\theta\left(t\right)-f_{n}\theta\left(-t\right)\right].
\end{equation}
The time domain screened potential $W_{0}\left(\boldsymbol{r},\boldsymbol{r}',t\right)$
is the potential at point $\boldsymbol{r}$ and time $t$ due to a
QP introduced at time $t=0$ at point $\boldsymbol{r}'$. Hence it
is composed of an instantaneous Coulomb term and a time dependent
polarization contribution: 
\begin{eqnarray}
W_{0}\left(\boldsymbol{r},\boldsymbol{r}',t\right) & = & u_{C}\left(\left|\boldsymbol{r}-\boldsymbol{r}'\right|\right)\delta\left(t\right)+W_{P}\left(\boldsymbol{r},\boldsymbol{r}',t\right).\label{eq:W(t)}
\end{eqnarray}
$W_{P}\left(\boldsymbol{r},\boldsymbol{r}',t\right)$ is the polarization
potential of the density perturbation due to the QP: 
\begin{equation}
W_{P}\left(\boldsymbol{r},\boldsymbol{r}',t\right)=\iint u_{C}\left(\left|\boldsymbol{r}-\boldsymbol{r}''\right|\right)\chi\left(\boldsymbol{r}'',\boldsymbol{r}''',t\right)u_{C}\left(\left|\boldsymbol{r}'''-\boldsymbol{r}'\right|\right)d\boldsymbol{r}''d\boldsymbol{r}''',\label{eq:WP}
\end{equation}
which is given in terms of the time-ordered reducible polarization
function $\chi\left(\boldsymbol{r},\boldsymbol{r}',t\right)$. Using
these definitions we write the self energy expectation value as a
sum of instantaneous and time-dependent contributions: 

\begin{equation}
\Sigma_{n}\left(t\right)=\Sigma_{n}^{X}\delta\left(t\right)+\Sigma_{n}^{P}\left(t\right).
\end{equation}
Here, the instantaneous contribution is 
\begin{equation}
\Sigma_{n}^{X}=-\iint\phi_{n}^{KS}\left(\boldsymbol{r}\right)u_{C}\left(\left|\boldsymbol{r}-\boldsymbol{r}'\right|\right)\rho^{KS}\left(\boldsymbol{r},\boldsymbol{r}'\right)\phi_{n}^{KS}\left(\boldsymbol{r}'\right)d\boldsymbol{r}d\boldsymbol{r}',\label{eq:Sigma_X}
\end{equation}
\emph{i.e.,} the expectation value of the exact exchange operator,
where 
\begin{equation}
\rho^{KS}\left(\boldsymbol{r},\boldsymbol{r}'\right)=-iG_{0}\left(\boldsymbol{r},\boldsymbol{r}',0^{-}\right)=\sum_{n}f_{n}\phi_{n}^{KS}\left(\boldsymbol{r}\right)\phi_{n}^{KS}\left(\boldsymbol{r}'\right)
\end{equation}
is the KS density matrix. Finally, the polarization self-energy is
given by the integral

\begin{equation}
\Sigma_{n}^{P}\left(t\right)=\iint\phi_{n}^{KS}\left(\boldsymbol{r}\right)iG_{0}\left(\boldsymbol{r},\boldsymbol{r}',t\right)W_{P}\left(\boldsymbol{r},\boldsymbol{r}',t^{+}\right)\phi_{n}^{KS}\left(\boldsymbol{r}'\right)d\boldsymbol{r}d\boldsymbol{r}'.\label{eq:SigmaP(t)}
\end{equation}
Despite the fact that the time-dependent formalism circumvents the
convolution appearing in the frequency-dependent domain, the numerical
evaluation of $\Sigma_{n}^{P}\left(t\right)$ is a significant challenge
with numerical effort typically scaling proportionally to $N_{e}^{4}$
or $N_{e}^{5}$.\cite{Deslippe2012,Nguyen2012,Pham2013} This is due
to the fact that $G_{0}\left(\boldsymbol{r},\boldsymbol{r}',t\right)$
involves all (occupied and unoccupied) KS orbitals and $W_{P}\left(\boldsymbol{r},\boldsymbol{r}',t\right)$
involves 6-dimensional integrals (Eq.~(\ref{eq:WP})) depending on
the reducible polarization function $\chi\left(\boldsymbol{r}'',\boldsymbol{r}''',t\right)$. 

\subsection{Stochastic $\boldsymbol{\text{\ensuremath{G_{0}W_{0}}}}$ }

We now explain how stochastic orbitals enable an efficient near-linear-scaling
calculation of $\Sigma_{n}\left(t\right)$.\cite{Neuhauser2014} The
calculation uses a real space 3D Cartesian grid with equally spaced
points $\boldsymbol{r}_{ijk}=\left(i\mathbf{\hat{x}}+j\mathbf{\hat{y}}+k\mathbf{\hat{z}}\right)h$,
where $i$, $j$ and $k$ are integers and $h$ is the grid spacing,
assumed for simplicity to be equal in the $x,y,z$ directions. The
application of the Kohn-Sham Hamiltonian $\hat{h}_{KS}$ onto any
function on the grid can be performed using Fast Fourier Transforms
in $N_{g}\log N_{g}$ scaling, where $N_{g}$ is the size of the grid. 

We now introduce a real stochastic orbital $\zeta\left(\boldsymbol{r}\right)$
on the grid assigning \emph{randomly} $+h^{-3/2}$ or $-h^{-3/2}$
with equal probability to $\zeta\left(\boldsymbol{r}\right)$ at each
grid point $\boldsymbol{r}$.\cite{Baer2013,Neuhauser2014a} The average
of the expectation value (expressed by $\left\langle \cdots\right\rangle _{\zeta}$)
of the projection $\left\langle \:\left|\zeta\left\rangle \right\langle \zeta\right|\:\right\rangle _{\zeta}$
is equal to the unit matrix, $\left\langle \,\left|\zeta\right\rangle \left\langle \zeta\right|\,\right\rangle _{\zeta}=\hat{\boldsymbol{I}}$,
resulting in a ``stochastic resolution of identity''.\cite{Hutchinson1990}
In practical calculations the expectation values, i.e., averages over
$\zeta$, are estimated using a finite sample of $N_{\zeta}$ random
states. According to the central limit theorem this average converges
to the expectation value as $N_{\zeta}\to\infty$ (for a discussion
of the convergence of the stochastic estimates see Sec.~\ref{sec:Results}).

Using the stochastic resolution of the identity any operator can be
represented as an average over a product of stochastic orbitals. For
example, for the KS Green's function:
\begin{equation}
iG_{0}\left(\boldsymbol{r},\boldsymbol{r}',t\right)=\left\langle \,\,\,\zeta\left(\boldsymbol{r}'\right)\zeta\left(\boldsymbol{r},t\right)\,\,\,\right\rangle _{\zeta},\label{eq:g0-stoch}
\end{equation}
where $\zeta\left(\boldsymbol{r}\right)=\left\langle \boldsymbol{r}|\zeta\right\rangle $
is the real random orbital and 
\begin{align}
\zeta\left(\boldsymbol{r},t\right) & =\left\langle \boldsymbol{r}|i\hat{G}_{0}\left(t\right)|\zeta\right\rangle \label{eq:zeta}\\
 & =\left\langle \boldsymbol{r}|e^{-i\hat{h}_{KS}t/\hbar}\left[\theta\left(t\right)-\theta_{\beta}\left(\mu-\hat{h}_{KS}\right)\right]|\zeta\right\rangle \nonumber 
\end{align}
 is the $G$-operated random orbital. Here, $\mu$ is the chemical
potential, $\theta\left(t\right)$ is the Heaviside function, and
$\theta_{\beta}\left(\varepsilon\right)=\frac{1}{2}\left[1+\text{erf}\left(\beta\varepsilon\right)\right]$
(in the limit $\beta\to\infty$, $\theta_{\beta}\left(\varepsilon\right)\to\theta\left(\beta\varepsilon\right)$).
The application of $i\hat{G}_{0}\left(t\right)$ on $\zeta$ in Eq.~(\ref{eq:g0-stoch})
is performed using a Chebyshev expansion (for applying $\theta_{\beta}\left(\mu-\hat{h}_{KS}\right)$)
and a split operator propagator for the time evolution, both taking
advantage of the sparsity of the KS Hamiltonian in the real-space
grid representation. The Chebyshev series includes a finite number
of terms $N_{C}\approx2\beta\Delta E$ where $\Delta E$ is the eigenvalue
range of the KS Hamiltonian $\hat{h}_{KS}$ and where $\beta$ is
large enough so that $\beta E_{g}\gg1$ where $E_{g}$ is the occupied-unoccupied
eigenvalue gap (see, e.g., Refs.~\citenum{Baer1997a,Baer1997b}).

The representation used in Eq.~(\ref{eq:g0-stoch}) decouples the
position-dependence on $\boldsymbol{r}$ and $\boldsymbol{r}^{\prime}$
and eliminates the need to represent $iG_{0}\left(\boldsymbol{r},\boldsymbol{r}',t\right)$
by all occupied and unoccupied orbitals. The polarization part of
the self-energy is recast as:

\[
\Sigma_{n}^{P}\left(t\right)=\left\langle \,\,\Sigma_{n\zeta}(t)\,\,\right\rangle _{\zeta},
\]

\begin{equation}
\Sigma_{n\zeta}^{P}\left(t\right)=\iint\phi_{n}^{KS}\left(\boldsymbol{r}\right)\zeta\left(\boldsymbol{r},t\right)W_{P}\left(\boldsymbol{r},\boldsymbol{r}',t\right)\phi_{n}^{KS}\left(\boldsymbol{r}'\right)\zeta\left(\boldsymbol{r}'\right)d\boldsymbol{r}d\boldsymbol{r}',\label{eq:Sigma_P_n}
\end{equation}
where $\zeta$ is the stochastic orbital used to characterize $G_{0}$.
Further simplifications are obtained by inserting yet another, independent,
real stochastic orbital $\xi\left(\boldsymbol{r}\right)$ using the
identity
\[
\phi_{n}^{KS}\left(\boldsymbol{r}\right)\zeta\left(\boldsymbol{r},t\right)W_{P}\left(\boldsymbol{r},\boldsymbol{r}',t\right)=\left\langle \int d\boldsymbol{r}''\phi_{n}^{KS}\left(\boldsymbol{r}''\right)\zeta\left(\boldsymbol{r}'',t\right)\xi\left(\boldsymbol{r}''\right)\xi\left(\boldsymbol{r}\right)W_{P}\left(\boldsymbol{r},\boldsymbol{r}',t\right)\right\rangle _{\xi},
\]
decoupling the two $t$-dependent functions. Therefore, the polarization
part of the self-energy becomes an average over a product of two time-dependent
stochastic functions $A_{n\zeta\xi}\left(t\right)$ and $B_{n\zeta\xi}\left(t\right)$:

\begin{equation}
\Sigma_{n\zeta}^{P}\left(t\right)=\left\langle \,\,A_{n\zeta\xi}\left(t\right)B_{n\zeta\xi}\left(t\right)\,\,\right\rangle _{\xi},\label{eq:StochSigmaP}
\end{equation}
where
\begin{equation}
A_{n\zeta\xi}\left(t\right)=\int\phi_{n}^{KS}\left(\boldsymbol{r}\right)\zeta\left(\boldsymbol{r},t\right)\xi\left(\boldsymbol{r}\right)d\boldsymbol{r}\label{eq:A_n-zeta-zeta}
\end{equation}
and
\begin{equation}
B_{n\zeta\xi}\left(t\right)=\iint\xi\left(\boldsymbol{r}\right)W_{P}\left(\boldsymbol{r},\boldsymbol{r}',t\right)\phi_{n}^{KS}\left(\boldsymbol{r}'\right)\zeta\left(\boldsymbol{r}'\right)d\boldsymbol{r}d\boldsymbol{r}'.\label{eq:B_n-zeta-zeta}
\end{equation}

Calculating $B_{n\zeta\xi}\left(t\right)$ is done efficiently using
the time-dependent Hartree (TDH) method equivalent to the popular
random phase approximation (RPA).\cite{Baer2004b} There is an important
caveat, however. The real-time formulation based on TDH provides a
description of the \emph{retarded} $W^{r}\left(\boldsymbol{r},\boldsymbol{r}',t\right)$
rather than the time-ordered $W_{P}\left(\boldsymbol{r},\boldsymbol{r}',t\right)$
needed in Eq.~\ref{eq:B_n-zeta-zeta}. Fortunately, in linear-response,
the two functions are simply related through the corresponding Fourier
transforms:\cite{Fetter1971}
\begin{equation}
\tilde{B}_{n\zeta\xi}\left(\omega\right)={\rm Re}\tilde{B}_{n\zeta\xi}^{r}\left(\omega\right)+i\,{\rm sign\left(\omega\right){\rm Im}}\tilde{B}_{n\zeta\xi}^{r}\left(\omega\right),\label{eq:retarded}
\end{equation}
where $\tilde{B}_{n\zeta\xi}^{r}$ is obtained with $W^{r}\left(\boldsymbol{r},\boldsymbol{r}',t\right)$.
Consequently, we first provide a formulation for $B_{n\zeta\xi}^{r}\left(t\right)$
and then, as mentioned, use Eq.~(\ref{eq:retarded}) to obtain the
corresponding time-ordered function $B_{n\zeta\xi}\left(t\right)$.

$B_{n\zeta\xi}^{r}\left(t\right)$ are obtained by combining the linear
response relation Eq.~(\ref{eq:WP}) (with $\chi^{r}$ replacing
$\chi)$ with the definition Eq. (\ref{eq:B_n-zeta-zeta}) yielding 

\begin{equation}
B_{n\zeta\xi}^{r}\left(t\right)=\iint\xi\left(\boldsymbol{r}\right)u_{C}\left(\left|\boldsymbol{r}-\boldsymbol{r}^{\prime}\right|\right)\Delta n_{n\zeta}^{r}\left(\boldsymbol{r}^{\prime},t\right)d\boldsymbol{r}d\boldsymbol{r}^{\prime},\label{eq:w-chi-v}
\end{equation}
which is calculated in near linear-scaling (rather than quadratic-scaling)
using Fast Fourier Transforms for the convolutions. Here, $\Delta n_{n\zeta}^{r}\left(\boldsymbol{r},t\right)$
is formally given by:

\begin{equation}
\Delta n_{n\zeta}^{r}\left(\boldsymbol{r},t\right)=\int\chi^{r}\left(\boldsymbol{r},\boldsymbol{r}^{\prime},t\right)v_{n\zeta}\left(\boldsymbol{r}^{\prime}\right)d\boldsymbol{r}^{\prime},
\end{equation}
 with

\begin{equation}
v_{n\zeta}\left(\boldsymbol{r}^{\prime}\right)=\int u_{C}\left(\left|\boldsymbol{r}^{\prime}-\boldsymbol{r}^{\prime\prime}\right|\right)\phi_{n}^{KS}\left(\boldsymbol{r}^{\prime\prime}\right)\zeta\left(\boldsymbol{r}^{\prime\prime}\right)d\boldsymbol{r}^{\prime\prime}.
\end{equation}
In practice, we calculate the density perturbation by taking $N_{\eta}$
stochastic orbitals $\bar{\eta}\left(\boldsymbol{r}\right)$ which
are projected on the occupied space using the Chebyshev expansion
of the operator $\theta_{\beta}\left(\mu-\hat{h}_{KS}\right)$, 
\begin{equation}
\eta=\theta_{\beta}\left(\mu-\hat{h}_{KS}\right)\bar{\eta}.\label{eq:eta_proj}
\end{equation}
Each orbital is then perturbed at time zero:
\begin{equation}
\eta_{\tau}\left(\boldsymbol{r},0\right)=e^{-iv_{n\zeta}\left(\boldsymbol{r}\right)\tau}\eta\left(\boldsymbol{r}\right)\label{eq:perturbation}
\end{equation}
where $\tau$ is a small-time parameter. In the RPA, the orbital is
now propagated in time by a TDH equation similar to the stochastic
time-dependent DFT:\cite{Gao2015}
\begin{equation}
i\frac{\partial}{\partial t}\eta_{\tau}\left(\boldsymbol{r},t\right)=\hat{h}_{KS}\,\eta_{\tau}\left(\boldsymbol{r},t\right)+\left(\int\frac{\Delta n_{n\zeta}^{r}\left(\boldsymbol{r}',t\right)}{\left|\boldsymbol{r}-\boldsymbol{r}'\right|}d\boldsymbol{r}'\right)\eta_{\tau}\left(\boldsymbol{r},t\right),\label{eq:t-propagation}
\end{equation}
where
\begin{equation}
\Delta n_{n\zeta}^{r}\left(\boldsymbol{r},t\right)=\frac{1}{\tau}\left\langle \left|\eta_{\tau}\left(\boldsymbol{r},t\right)\right|^{2}-\left|\eta_{\tau=0}\left(\boldsymbol{r},t\right)\right|^{2}\right\rangle _{\eta}.\label{eq:delta n(r,t)}
\end{equation}

From $\Delta n_{n\zeta}^{r}\left(\boldsymbol{r},t\right)$ we then
evaluate $B_{n\zeta\xi}^{r}\left(t\right)$ via Eq. (\ref{eq:eta_proj}),
and then Fourier transform the coefficients from time to frequency
and back via Eq. (\ref{eq:retarded}) to yield the required $B_{n\zeta\xi}\left(t\right).$ 

Finally, the exchange part of the self energy is simplified, by replacing
the 6-dimensional integral in Eq.~\ref{eq:Sigma_X} by two $3$-dimensional
integrals involving projected occupied orbitals
\begin{equation}
\Sigma_{n}^{X}=-\left\langle \int\phi_{n}^{KS}\left(\boldsymbol{r}\right)\eta(\boldsymbol{r})v_{\eta}^{{\rm aux}}\left(\boldsymbol{r}\right)d\boldsymbol{r}\right\rangle _{\eta},\label{eq:sSigX}
\end{equation}
where the auxiliary potential is 
\begin{equation}
v_{\eta}^{{\rm aux}}\left(\boldsymbol{r}\right)=\int u_{C}\left(\left|\boldsymbol{r}-\boldsymbol{r}'\right|\right)\eta(\boldsymbol{r})\phi\left(\boldsymbol{r}'\right)d\boldsymbol{r}'.\label{eq:Vaux}
\end{equation}
Note that we are allowed to use the same projected states $\eta$
obtained from Eq.~(\ref{eq:eta_proj}) also for calculating the exchange
part, which is therefore obtained automatically as a byproduct of
the polarization self-energy with essentially no extra cost.

\subsection{The algorithm }

We summarize the procedure above by the following algorithm for computing
the sGW QP energies: 
\begin{enumerate}
\item Generate a stochastic orbital $\zeta\left(\boldsymbol{r}\right)$
and $N_{\xi}$ stochastic orbitals $\xi\left(\boldsymbol{r}\right)$.
Use Eq.~(\ref{eq:zeta}) to generate the projected time-dependent
orbital $\zeta\left(\boldsymbol{r},t\right)$.
\item Generate the set of $N_{\xi}$ time-dependent function $A_{n\zeta\xi}\left(t\right)$
from Eq.~(\ref{eq:A_n-zeta-zeta}) using $\xi\left(\boldsymbol{r}\right)$
and $\zeta\left(\boldsymbol{r},t\right)$.
\item Generate $N_{\eta}$ independent stochastic orbitals, project each
of them to the occupied subspace according Eq.~(\ref{eq:eta_proj}),
obtaining the projected $N_{\eta}$ functions $\eta(\boldsymbol{r})$
from which $\Sigma_{n}^{X}$ is computed using Eqs.~(\ref{eq:sSigX})-(\ref{eq:Vaux}).
\item Then use the same $N_{\eta}$ projected stochastic functions $\eta(\boldsymbol{r})$
together with $\zeta\left(\boldsymbol{r}\right)$ and the set of $\xi\left(\boldsymbol{r}\right)$
to generate $B_{n\zeta\xi}^{r}\left(t\right)$ using Eqs.~(\ref{eq:B_n-zeta-zeta})-(\ref{eq:delta n(r,t)}),
where $n_{n\zeta}^{r}\left(\boldsymbol{r},t\right)$ is obtained as
an average over $\eta$.
\item Fourier transform $B_{n\zeta\xi}^{r}\left(t\right)\to\tilde{B}_{n\zeta\xi}^{r}\left(\omega\right)$
and convert to the time-ordered quantity $\tilde{B}_{n\zeta\xi}\left(\omega\right)$
using Eq.~(\ref{eq:retarded}). Fourier transform back $\tilde{B}_{n\zeta\xi}^{TO}\left(\omega\right)\to B_{n\zeta\xi}^{TO}\left(t\right)$
and calculate, by averaging on $\xi$, the polarization self-energy
$\Sigma_{n\zeta}^{P}\left(t\right)$ using Eq.~(\ref{eq:B_n-zeta-zeta}).
\item Repeat steps 1-5 $N_{\zeta}$ times, averaging $\Sigma_{n}^{P}\left(t\right)=\frac{1}{N_{\zeta}}\sum_{\zeta}\big(\Sigma_{n\zeta}^{P}\left(t\right)\big)$
and similarly averaging $\Sigma_{n}^{X}$ .
\item Fourier transform $\Sigma_{n}^{P}\left(t\right)\to\tilde{\Sigma}_{n}^{P}\left(\omega\right)$
and using this function estimate the QP energy $\varepsilon_{n}^{QP}$
by solving Eq.(\ref{eq:perturbativeQPE}) self-consistently
\end{enumerate}
In practice, the stochastic error is then estimated by dividing the
set of $N_{\zeta}$ calculations to e.g., 100 subsets (in each of
which we use $\frac{N_{\zeta}}{100}$ stochastic orbitals) and then
estimating the error based on the values of $\varepsilon_{n}^{QP}$
from each of the 100 subsets.

\begin{figure}
\includegraphics[width=0.475\textwidth]{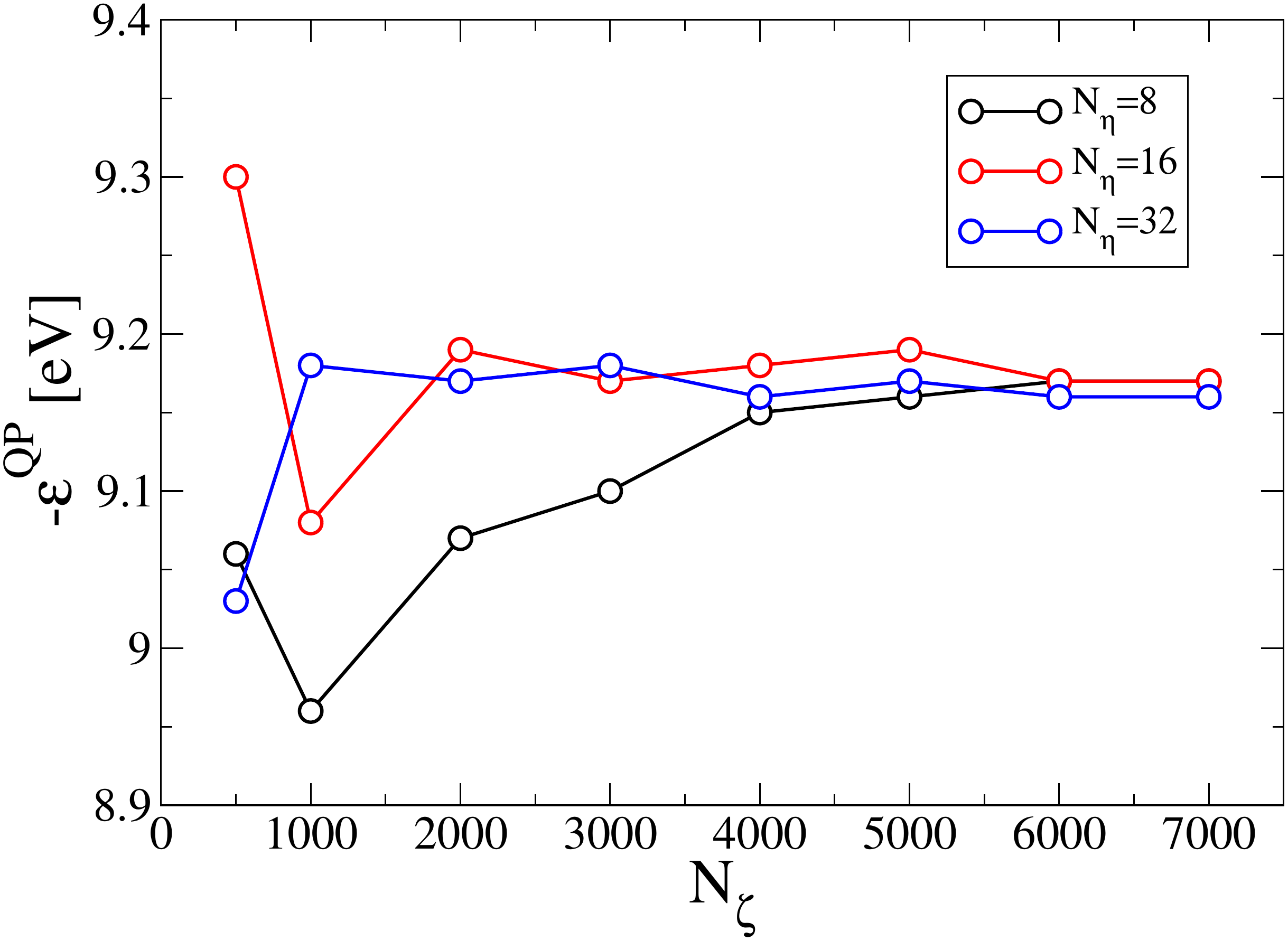}~~\includegraphics[width=0.475\textwidth]{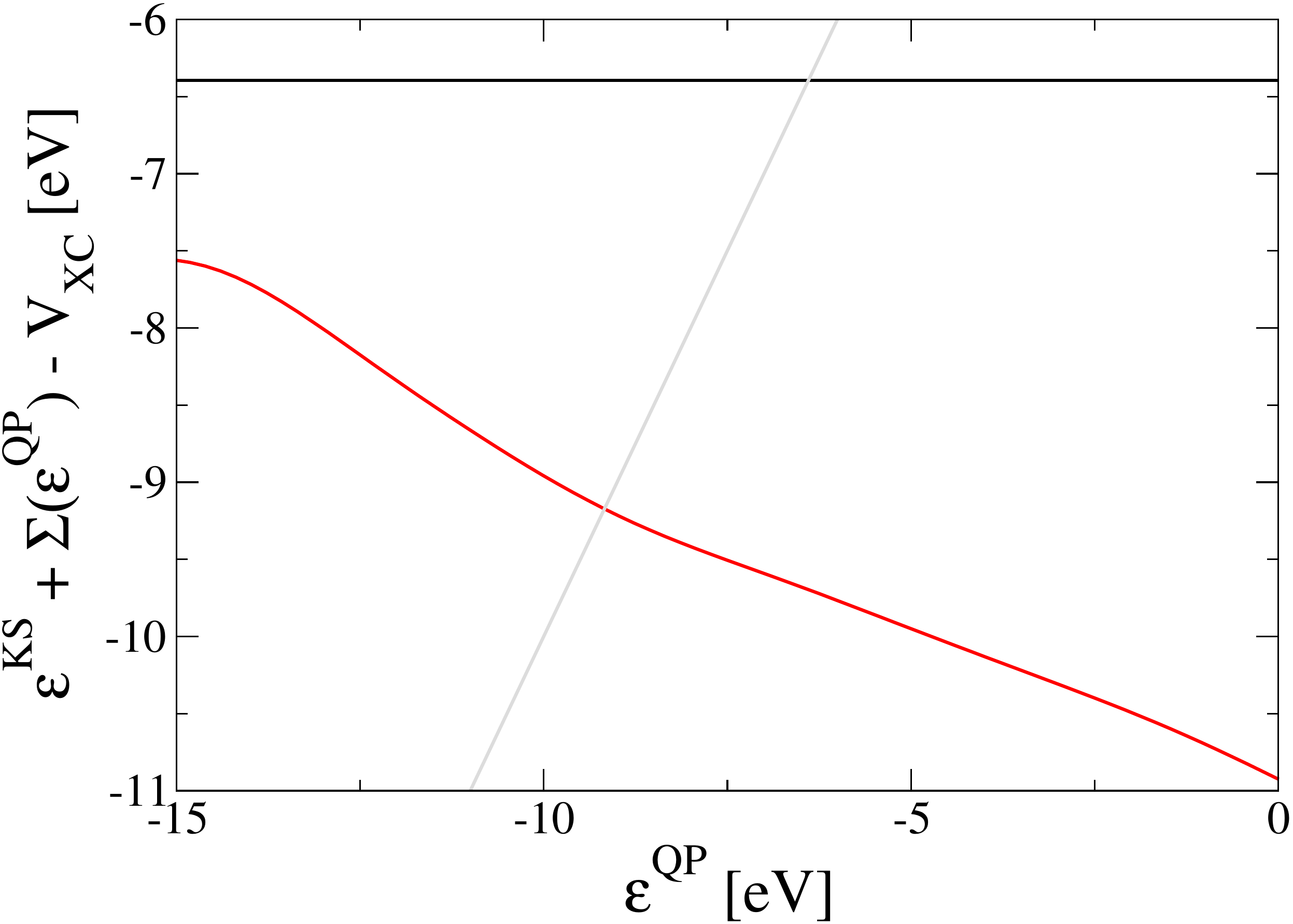}

\caption{\label{fig:Benzene-convergence}Left panel: Convergence of the sGW
estimate of the QP hole energy for a benzene molecule as a function
of $N_{\zeta}$ for different values of $N_{\eta}$. Right: A graphic
representation of the self-consistent solution of Eq.~(\ref{eq:perturbativeQPE})
for $-IP=\varepsilon^{QP}$ of benzene. The solid red line represent
the right hand side of Eq.~(\ref{eq:perturbativeQPE}). The intersect
with the solid gray line represents the self-consistent solution.
For reference, we also depict $\varepsilon^{KS}$ (solid black line).}
\end{figure}

\section{\label{sec:Results}Results }

We now evaluate the performance of sGW by application to a set of
10 small enough molecules for which reliable deterministic calculations
and experimental vertical ionization energies are available. The sGW
calculation is based on the local density approximation, denoted henceforth
as $\varepsilon_{\text{@LDA}}^{\text{sGW}}$ and implemented on a
Fourier real-space grid using Troullier-Martins pseudopotentials \cite{Troullier1991}
and the technique for screening periodic charge images of Ref.~\citenum{}Martyna1999.
For all molecules experimental geometries were used, taken from the
NIST database.\cite{NISTDataBase} 

The sGW estimate of $\varepsilon_{\text{@LDA}}^{\text{sGW}}$ is governed
by convergence of multiple parameters. The grid spacing was determined
in the preparatory DFT step by requiring convergence of the LDA eigenvalues
to better than 1meV (our LDA eigenvalues deviate by 0.03eV or less
from those obtained by the QuantumEspresso program using the same
pseudopotentials). For all molecules we chose the inverse temperature
parameter as $\beta=200E_{h}^{-1}$ from which the Chebyshev expansion
length $N_{C}$ was derived to be between 18,000 and 19,000 (see discussion
appearing below Eq.~(\ref{eq:zeta})). The time propagation is performed
using a discretized time-step of $\Delta t=0.05E_{h}^{-1}\hbar$ for
both the Green's function calculation as well as the RPA screening,
we checked that this leads to QP energies converged to within less
than 0.02eV. 

Other parameters only negligibly influence the result. Specifically,
the strength of the perturbation was controlled by the parameter $\tau$
(see Eq.~(\ref{eq:perturbation})); changing its value between $0.01$
to $0.0001\;E_{h}^{-1}\hbar$ influences the QP energy by less than
$0.001$eV. In practice we employ $\tau=0.001\;E_{h}^{-1}\hbar$.
Furthermore, we used $N_{\xi}=100$ and ascertained that increasing
this value to $200$ causes changes in the QP energies smaller than
0.01 eV. 

The most influential parameters are $N_{\eta}$, the number of stochastic
states $\eta$ used for the RPA screening calculation, and $N_{\zeta}$
used for representing the Green's function. In the left panel of Fig.~\ref{fig:Benzene-convergence}
the convergence of the QP energy for a benzene molecule is illustrated
as a function of $N_{\zeta}$ for several values of $N_{\eta}$. Evidently,
for this molecule, $N_{\zeta}=6000$ and $N_{\eta}=8$ are sufficient
to converge the QP energy with a statistical error of $\pm0.03\,\text{eV}$.
Note that as $N_{\eta}$ increases the convergence towards the final
QP value is reached after a smaller number of $N_{\zeta}$ stochastic
orbitals. 

When transforming from the time to the frequency domain we use a Gaussian
damping factor, $\tilde{B}_{n\zeta\xi}^{r}\left(\omega\right)=\int_{0}^{T}dt\,e^{i\omega t}B_{n\zeta\xi}^{r}\left(t\right)\times e^{-\left(\gamma t\right)^{2}/2}$,
where $\gamma=0.04E_{h}\hbar^{-1}$ and $T\approx4/\gamma=100\hbar E_{h}^{-1}$
are enough to yield QP energies converged to within 0.01~eV. Note
that a value of $N_{\eta}=8$ is sufficient for a stable and accurate
time propagation up to $T=100\hbar E_{h}^{-1}$ but when longer times
$T$ are used, $N_{\eta}$ must be increased accordingly due to an
instability in stochastic TDDFT time propagation.\cite{Rabani2015} 

The right panel of Fig.~\ref{fig:Benzene-convergence} provides a
graphic representation of the self-consistent solution of Eq.~(\ref{eq:perturbativeQPE})
as the intersect between $\varepsilon^{QP}$ and $\varepsilon^{KS}+\Sigma\left(\varepsilon^{QP}\right)-V_{XC}$
. Note that even though the stochastic calculation has by its nature
fluctuations, the energy dependence of $\Sigma\left(\varepsilon^{QP}\right)$
is smooth. 

\begin{figure}
\includegraphics[width=0.7\textwidth]{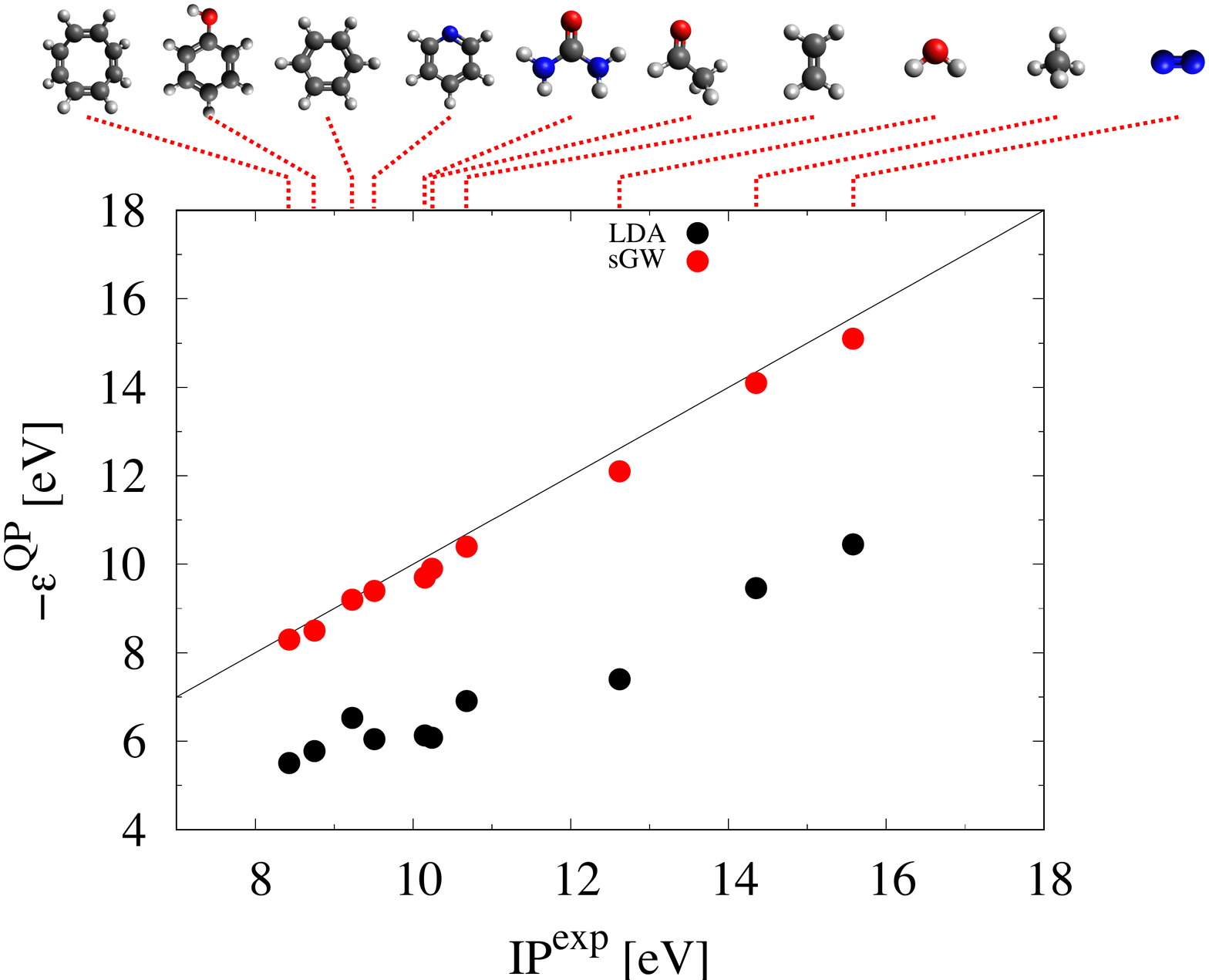}

\caption{\label{fig:Ionization-potential} Ionization potentials as predicted
by various calculations for the set of molecules listed in Table~\ref{tab:GW-benchmark}
are plotted against experimental values (note that the sGW statistical
error bars are smaller than the corresponding symbol sizes). Each
molecule is depicted above the graph and dotted red line points to
its experimental ionization potential on the horizontal axis. The
sketches of the individual molecules use black, white, blue and red
spheres to indicate positions of C, H, N and O atoms respectively.
LDA results that served as a starting point for the calculations are
shown by black circles. $\text{\ensuremath{G_{0}}}\text{\ensuremath{W_{0}}}$
results are given by filled red circles. The black line represents
the one-to-one correspondence to experimental values.}
\end{figure}

The sGW estimated vertical ionization energies $\varepsilon_{\text{@LDA}}^{\text{sGW}}$
were converged with respect to all parameters described above and
especially, grid-size and number of stochastic orbitals $N_{\zeta}$.
Hence, they should be compared to deterministic GW results which are
of a complete basis set quality at the GW@LDA level, denoted $\varepsilon_{\text{@LDA}}^{\text{EXTRA}}$,
extrapolated to the complete basis set limit. These results were based
on the GW@PBE extrapolated results $\varepsilon_{\text{@PBE}}^{\text{EXTRA}}$
calculated under the FHI-aims code \cite{Blum2009,Ren2012a} as given
in Ref.~\citenum{vanSetten2015}, which were then augmented for LDA
based energies using the relation: 

\begin{equation}
\varepsilon_{\text{@LDA}}^{\text{EXTRA}}\equiv\varepsilon_{\text{@PBE}}^{\text{EXTRA}}+(\varepsilon_{\text{@LDA}}^{\text{RI}}-\varepsilon_{\text{@PBE}}^{\text{RI}}),\label{eq:dGW@LDA}
\end{equation}
where $\varepsilon_{\text{@LDA}}^{\text{RI}}-\varepsilon_{\text{@PBE}}^{\text{RI}}$
is an estimate of the difference between PBE and LDA based GW results
(typically a very small energy in the range 0.01-0.08 eV). $\varepsilon_{\text{@LDA}}^{\text{RI}}$
and $\varepsilon_{\text{@PBE}}^{\text{RI}}$ are the GW-TURBOMOLE
\cite{vanSetten2013} energies calculated using the def2-QZVP basis-set
and the resolution-of-identity (RI) approximation. The switch between
FHI-aims code and GW-TURBOMOLE codes is not expected to pose a problem
since both give almost identical excitation energies \cite{vanSetten2015}.
We have also ascertained, using several tests on small molecules,
that $\varepsilon_{\text{@LDA}}^{\text{RI}}-\varepsilon_{\text{@PBE}}^{\text{RI}}$
is quite independent of the RI approximation (even though RI does
affect the separate values of each energy). 

\begin{table}
\begin{tabular}{|c|c|c|c|c|c|c||c|c|}
\hline 
\multirow{2}{*}{{\scriptsize{}System}} & \multirow{2}{*}{{\scriptsize{}Exp.}} & \multirow{2}{*}{{\scriptsize{}$\varepsilon_{\text{@PBE}}^{\text{EXTRA}}$}} & \multirow{2}{*}{{\scriptsize{}$\varepsilon_{\text{@LDA}}^{\text{RI}}-\varepsilon_{\text{@PBE}}^{\text{RI}}$}} & \multirow{2}{*}{{\scriptsize{}$\varepsilon{}_{\text{@LDA}}^{\text{EXTRA}}$}} & \multirow{2}{*}{{\scriptsize{}$\varepsilon_{\text{@LDA}}^{\text{sGW}}$}} & \multirow{2}{*}{{\scriptsize{}Diff}} & \multirow{2}{*}{{\scriptsize{}$h/a_{0}$}} & \multirow{2}{*}{{\scriptsize{}$N_{\zeta}$}}\tabularnewline
 &  &  &  &  &  &  &  & \tabularnewline
\hline 
{\scriptsize{}benzene} & {\scriptsize{}9.23} & {\scriptsize{}9.10(0.01)} & {\scriptsize{}0.03} & \textbf{\scriptsize{}9.13} & \textbf{\scriptsize{}9.17(0.03)} & {\scriptsize{}0.04} & {\scriptsize{}0.30} & {\scriptsize{}6000}\tabularnewline
\hline 
{\scriptsize{}cyclooctatetraene} & {\scriptsize{}8.43} & {\scriptsize{}8.18(0.02)} & {\scriptsize{}0.02} & \textbf{\scriptsize{}8.20} & \textbf{\scriptsize{}8.33(0.03)} & {\scriptsize{}0.13} & {\scriptsize{}0.35} & {\scriptsize{}6000}\tabularnewline
\hline 
{\scriptsize{}acetaldehyde} & {\scriptsize{}10.20} & {\scriptsize{}9.66(0.03)} & {\scriptsize{}0.08} & \textbf{\scriptsize{}9.74} & \textbf{\scriptsize{}9.90(0.06)} & {\scriptsize{}0.16} & {\scriptsize{}0.30} & {\scriptsize{}8000}\tabularnewline
\hline 
{\scriptsize{}water} & {\scriptsize{}12.60} & {\scriptsize{}12.05(0.03)} & {\scriptsize{}0.08} & \textbf{\scriptsize{}12.13} & \textbf{\scriptsize{}12.10(0.02)} & {\scriptsize{}-0.04} & {\scriptsize{}0.25} & {\scriptsize{}8000}\tabularnewline
\hline 
{\scriptsize{}phenol} & {\scriptsize{}8.75} & {\scriptsize{}8.51(0.01)} & {\scriptsize{}0.05} & \textbf{\scriptsize{}8.56} & \textbf{\scriptsize{}8.61(0.03)} & {\scriptsize{}0.05} & {\scriptsize{}0.35} & {\scriptsize{}9000}\tabularnewline
\hline 
{\scriptsize{}urea} & {\scriptsize{}10.15} & {\scriptsize{}9.46(0.02)} & {\scriptsize{}0.12} & \textbf{\scriptsize{}9.58} & \textbf{\scriptsize{}9.65(0.05)} & {\scriptsize{}0.07} & {\scriptsize{}0.30} & {\scriptsize{}11000}\tabularnewline
\hline 
{\scriptsize{}methane} & {\scriptsize{}14.40} & {\scriptsize{}14.00(0.06)} & {\scriptsize{}0.03} & \textbf{\scriptsize{}14.03} & \textbf{\scriptsize{}14.09(0.01)} & {\scriptsize{}0.06} & {\scriptsize{}0.40} & {\scriptsize{}10000}\tabularnewline
\hline 
{\scriptsize{}nitrogen} & {\scriptsize{}15.60} & {\scriptsize{}15.05(0.04)} & {\scriptsize{}0.11} & \textbf{\scriptsize{}15.16} & \textbf{\scriptsize{}15.05(0.06)} & {\scriptsize{}-0.11} & {\scriptsize{}0.35} & {\scriptsize{}7000}\tabularnewline
\hline 
{\scriptsize{}ethylene} & {\scriptsize{}10.70} & {\scriptsize{}10.40(0.03)} & {\scriptsize{}0.03} & \textbf{\scriptsize{}10.43} & \textbf{\scriptsize{}10.40(0.06)} & {\scriptsize{}-0.03} & {\scriptsize{}0.35} & {\scriptsize{}12000}\tabularnewline
\hline 
{\scriptsize{}pyridine} & {\scriptsize{}9.50} & {\scriptsize{}9.17(0.01)} & {\scriptsize{}0.06} & \textbf{\scriptsize{}9.23} & \textbf{\scriptsize{}9.42(0.04)} & {\scriptsize{}0.19} & {\scriptsize{}0.35} & {\scriptsize{}7000}\tabularnewline
\hline 
\multicolumn{1}{c}{} & \multicolumn{1}{c}{} & \multicolumn{1}{c}{} & \multicolumn{1}{c}{} &  & \multicolumn{1}{r|}{{\scriptsize{}Mean:}} & \multicolumn{1}{c|}{{\scriptsize{}0.05}} & \multicolumn{1}{c}{} & \multicolumn{1}{c}{}\tabularnewline
\cline{6-7} 
\multicolumn{1}{c}{} & \multicolumn{1}{c}{} & \multicolumn{1}{c}{} & \multicolumn{1}{c}{} &  & \multicolumn{1}{r|}{{\scriptsize{}Mean Abs:}} & \multicolumn{1}{c|}{{\scriptsize{}0.09}} & \multicolumn{1}{c}{} & \multicolumn{1}{c}{}\tabularnewline
\cline{6-7} 
\end{tabular}

\caption{\label{tab:GW-benchmark} Vertical ionization energies (eV) for the
indicated molecules. The complete-basis-set-limit-extrapolated GW@PBE
result, $\varepsilon_{\text{@PBE}}^{\text{EXTRA}}$ (with extrapolation
uncertainties in parenthesis) is taken from Ref.~\citenum{vanSetten2015}.
$\varepsilon_{\text{@LDA}}^{\text{RI}}-\varepsilon_{\text{@PBE}}^{\text{RI}}$is
the estimated GW@LDA to GW@PBE difference, calculated using GW-TURBOMOLE
\cite{vanSetten2013} within the RI approximation and the def2-QZVP
basis-set. The extrapolated GW@LDA results $\varepsilon{}_{\text{@LDA}}^{\text{EXTRA}}$
represent our estimate of the fully converged GW@LDA energies (given
in Eq.~\ref{eq:dGW@LDA}) which are used to benchmark the sGW@LDA
energies $\varepsilon{}_{\text{@LDA}}^{\text{sGW}}$ (with statistical
uncertainties given in parenthesis). For each molecule, the grid spacing
$h$ and the number of stochastic orbitals $N_{\zeta}$ required for
producing converged sGW to the indicated accuracy are given in the
table. }
\end{table}

In Table~\ref{tab:GW-benchmark} we compare the GW and sGW LDA-based
vertical ionization energies, showing a high level of agreement, with
mean and absolute deviations of $0.05\,\text{eV}$ and $0.09$eV respectively,
typically of the order of the given uncertainties in the deterministic
and the stochastic calculations. 

We also note that both these values are also in good overall agreement
with experimental values, as seen in Fig.~\ref{fig:Ionization-potential},
although both results (stochastic or deterministic) generally underestimate
the experiment by 0.1-0.5 eV. This is primarily due to the known limitations
of the $\text{G}_{0}\text{W}_{0}$ approach, which can be improved
using self consistent-GW.\cite{Shishkin2007,Caruso2012,Kaplan2016,vlcek2017self}

\section{\label{sec:Summary}Conclusions}

In conclusion, we reviewed in detail the sGW method and its algorithmic
implementation. The sGW exhibits a near-linear scaling with system
size complexity \cite{Neuhauser2014} and hence for large systems
it is expected to be much faster relative to the deterministic basis-set
implementations having quartic or quintic \cite{vanSetten2013} asymptotic
scaling. Therefore, comparison of sGW estimations with those of deterministic
GW can only be made on relatively small molecules and here we selected
a set of 10 such molecules having $N_{e}=10-50$ electrons. For this
set, the execution time of sGW was larger than that of deterministic
GW codes and we estimate that the crossover would occur for molecules
with $N_{e}\approx200$. For the selected set of molecules, sGW and
deterministic GW predicted vertical ionization energies which were
very close, with maximal deviation smaller than 0.2 eV and average
and absolute deviations of 0.05eV and 0.1eV.
\begin{acknowledgement}
This work was supported by the Israel Science Foundation \textendash{}
FIRST Program (Grant No. 1700/14) and the Center for Computational
Study of Excited-State Phenomena in Energy Materials at the Lawrence
Berkeley National Laboratory, which is funded by the U.S. Department
of Energy, Office of Science, Basic Energy Sciences, Materials Sciences
and Engineering Division under Contract No. DE-AC02-05CH11231, as
part of the Computational Materials Sciences Program. Some of the
calculations were performed as part of the XSEDE computational project
TG-CHE160092.
\end{acknowledgement}
\addcontentsline{toc}{section}{\refname}
\providecommand{\latin}[1]{#1}
\makeatletter
\providecommand{\doi}
  {\begingroup\let\do\@makeother\dospecials
  \catcode`\{=1 \catcode`\}=2\doi@aux}
\providecommand{\doi@aux}[1]{\endgroup\texttt{#1}}
\makeatother
\providecommand*\mcitethebibliography{\thebibliography}
\csname @ifundefined\endcsname{endmcitethebibliography}
  {\let\endmcitethebibliography\endthebibliography}{}


\providecommand{\latin}[1]{#1}
\makeatletter
\providecommand{\doi}
  {\begingroup\let\do\@makeother\dospecials
  \catcode`\{=1 \catcode`\}=2\doi@aux}
\providecommand{\doi@aux}[1]{\endgroup\texttt{#1}}
\makeatother
\providecommand*\mcitethebibliography{\thebibliography}
\csname @ifundefined\endcsname{endmcitethebibliography}
  {\let\endmcitethebibliography\endthebibliography}{}
\begin{mcitethebibliography}{48}
\providecommand*\natexlab[1]{#1}
\providecommand*\mciteSetBstSublistMode[1]{}
\providecommand*\mciteSetBstMaxWidthForm[2]{}
\providecommand*\mciteBstWouldAddEndPuncttrue
  {\def\EndOfBibitem{\unskip.}}
\providecommand*\mciteBstWouldAddEndPunctfalse
  {\let\EndOfBibitem\relax}
\providecommand*\mciteSetBstMidEndSepPunct[3]{}
\providecommand*\mciteSetBstSublistLabelBeginEnd[3]{}
\providecommand*\EndOfBibitem{}
\mciteSetBstSublistMode{f}
\mciteSetBstMaxWidthForm{subitem}{(\alph{mcitesubitemcount})}
\mciteSetBstSublistLabelBeginEnd
  {\mcitemaxwidthsubitemform\space}
  {\relax}
  {\relax}

\bibitem[Hohenberg and Kohn(1964)Hohenberg, and Kohn]{Hohenberg1964}
Hohenberg,~P.; Kohn,~W. Inhomogeneous Electron Gas. \emph{Phys. Rev.}
  \textbf{1964}, \emph{136}, B864\relax
\mciteBstWouldAddEndPuncttrue
\mciteSetBstMidEndSepPunct{\mcitedefaultmidpunct}
{\mcitedefaultendpunct}{\mcitedefaultseppunct}\relax
\EndOfBibitem
\bibitem[Kohn and Sham(1965)Kohn, and Sham]{Kohn1965}
Kohn,~W.; Sham,~L.~J. Self-Consistent Equations Including Exchange and
  Correlation Effects. \emph{Phys. Rev.} \textbf{1965}, \emph{140}, A1133\relax
\mciteBstWouldAddEndPuncttrue
\mciteSetBstMidEndSepPunct{\mcitedefaultmidpunct}
{\mcitedefaultendpunct}{\mcitedefaultseppunct}\relax
\EndOfBibitem
\bibitem[VandeVondele \latin{et~al.}(2012)VandeVondele, Borstnik, and
  Hutter]{VandeVondeleBorstnikHutter2012}
VandeVondele,~J.; Borstnik,~U.; Hutter,~J.~r. Linear scaling self-consistent
  field calculations with millions of atoms in the condensed phase. \emph{J.
  Chem. Theory Comput.} \textbf{2012}, \emph{8}, 3565--3573\relax
\mciteBstWouldAddEndPuncttrue
\mciteSetBstMidEndSepPunct{\mcitedefaultmidpunct}
{\mcitedefaultendpunct}{\mcitedefaultseppunct}\relax
\EndOfBibitem
\bibitem[Hybertsen and Louie(1985)Hybertsen, and Louie]{Hybertsen1985}
Hybertsen,~M.~S.; Louie,~S.~G. First-principles theory of quasiparticles:
  calculation of band gaps in semiconductors and insulators. \emph{Phys. Rev.
  Lett.} \textbf{1985}, \emph{55}, 1418\relax
\mciteBstWouldAddEndPuncttrue
\mciteSetBstMidEndSepPunct{\mcitedefaultmidpunct}
{\mcitedefaultendpunct}{\mcitedefaultseppunct}\relax
\EndOfBibitem
\bibitem[Steinbeck \latin{et~al.}(1999)Steinbeck, Rubio, Reining, Torrent,
  White, and Godby]{Steinbeck1999}
Steinbeck,~L.; Rubio,~A.; Reining,~L.; Torrent,~M.; White,~I.; Godby,~R.
  Enhancements to the GW space-time method. \emph{Comput. Phys. Commun.}
  \textbf{1999}, \emph{125}, 05--118\relax
\mciteBstWouldAddEndPuncttrue
\mciteSetBstMidEndSepPunct{\mcitedefaultmidpunct}
{\mcitedefaultendpunct}{\mcitedefaultseppunct}\relax
\EndOfBibitem
\bibitem[Shishkin and Kresse(2007)Shishkin, and Kresse]{Shishkin2007}
Shishkin,~M.; Kresse,~G. Self-consistent GW calculations for semiconductors and
  insulators. \emph{Phys. Rev. B} \textbf{2007}, \emph{75}, 235102\relax
\mciteBstWouldAddEndPuncttrue
\mciteSetBstMidEndSepPunct{\mcitedefaultmidpunct}
{\mcitedefaultendpunct}{\mcitedefaultseppunct}\relax
\EndOfBibitem
\bibitem[Rostgaard \latin{et~al.}(2010)Rostgaard, Jacobsen, and
  Thygesen]{Rostgaard2010}
Rostgaard,~C.; Jacobsen,~K.~W.; Thygesen,~K.~S. Fully self-consistent GW
  calculations for molecules. \emph{Phys. Rev. B} \textbf{2010}, \emph{81},
  085103\relax
\mciteBstWouldAddEndPuncttrue
\mciteSetBstMidEndSepPunct{\mcitedefaultmidpunct}
{\mcitedefaultendpunct}{\mcitedefaultseppunct}\relax
\EndOfBibitem
\bibitem[Foerster \latin{et~al.}(2011)Foerster, Koval, and
  S{\'a}nchez-Portal]{Foerster2011}
Foerster,~D.; Koval,~P.; S{\'a}nchez-Portal,~D. An O (N3) implementation of
  Hedin's GW approximation for molecules. \emph{J. Chem. Phys.} \textbf{2011},
  \emph{135}, 074105\relax
\mciteBstWouldAddEndPuncttrue
\mciteSetBstMidEndSepPunct{\mcitedefaultmidpunct}
{\mcitedefaultendpunct}{\mcitedefaultseppunct}\relax
\EndOfBibitem
\bibitem[Faber \latin{et~al.}(2011)Faber, Attaccalite, Olevano, Runge, and
  Blase]{Faber2011}
Faber,~C.; Attaccalite,~C.; Olevano,~V.; Runge,~E.; Blase,~X. First-principles
  GW calculations for {DNA} and RNA nucleobases. \emph{Phys. Rev. B}
  \textbf{2011}, \emph{83}, 115123\relax
\mciteBstWouldAddEndPuncttrue
\mciteSetBstMidEndSepPunct{\mcitedefaultmidpunct}
{\mcitedefaultendpunct}{\mcitedefaultseppunct}\relax
\EndOfBibitem
\bibitem[Blase \latin{et~al.}(2011)Blase, Attaccalite, and Olevano]{Blase2011a}
Blase,~X.; Attaccalite,~C.; Olevano,~V. First-principles GW calculations for
  fullerenes, porphyrins, phtalocyanine, and other molecules of interest for
  organic photovoltaic applications. \emph{Phys. Rev. B} \textbf{2011},
  \emph{83}, 115103\relax
\mciteBstWouldAddEndPuncttrue
\mciteSetBstMidEndSepPunct{\mcitedefaultmidpunct}
{\mcitedefaultendpunct}{\mcitedefaultseppunct}\relax
\EndOfBibitem
\bibitem[Deslippe \latin{et~al.}(2012)Deslippe, Samsonidze, Strubbe, Jain,
  Cohen, and Louie]{Deslippe2012}
Deslippe,~J.; Samsonidze,~G.; Strubbe,~D.~A.; Jain,~M.; Cohen,~M.~L.;
  Louie,~S.~G. BerkeleyGW: A massively parallel computer package for the
  calculation of the quasiparticle and optical properties of materials and
  nanostructures. \emph{Comput. Phys. Commun.} \textbf{2012}, \emph{183}, 1269
  -- 1289\relax
\mciteBstWouldAddEndPuncttrue
\mciteSetBstMidEndSepPunct{\mcitedefaultmidpunct}
{\mcitedefaultendpunct}{\mcitedefaultseppunct}\relax
\EndOfBibitem
\bibitem[Marom \latin{et~al.}(2012)Marom, Caruso, Ren, Hofmann,
  K{\"o}rzd{\"o}rfer, Chelikowsky, Rubio, Scheffler, and Rinke]{Marom2012}
Marom,~N.; Caruso,~F.; Ren,~X.; Hofmann,~O.~T.; K{\"o}rzd{\"o}rfer,~T.;
  Chelikowsky,~J.~R.; Rubio,~A.; Scheffler,~M.; Rinke,~P. Benchmark of GW
  methods for azabenzenes. \emph{Phys. Rev. B} \textbf{2012}, \emph{86},
  245127\relax
\mciteBstWouldAddEndPuncttrue
\mciteSetBstMidEndSepPunct{\mcitedefaultmidpunct}
{\mcitedefaultendpunct}{\mcitedefaultseppunct}\relax
\EndOfBibitem
\bibitem[Caruso \latin{et~al.}(2012)Caruso, Rinke, Ren, Scheffler, and
  Rubio]{Caruso2012}
Caruso,~F.; Rinke,~P.; Ren,~X.; Scheffler,~M.; Rubio,~A. Unified description of
  ground and excited states of finite systems: The self-consistent GW approach.
  \emph{Phys. Rev. B} \textbf{2012}, \emph{86}, 081102\relax
\mciteBstWouldAddEndPuncttrue
\mciteSetBstMidEndSepPunct{\mcitedefaultmidpunct}
{\mcitedefaultendpunct}{\mcitedefaultseppunct}\relax
\EndOfBibitem
\bibitem[van Setten \latin{et~al.}(2013)van Setten, Weigend, and
  Evers]{vanSetten2013}
van Setten,~M.; Weigend,~F.; Evers,~F. The GW-Method for Quantum Chemistry
  Applications: Theory and Implementation. \emph{J. Chem. Theory Comput.}
  \textbf{2013}, \emph{9}, 232--246\relax
\mciteBstWouldAddEndPuncttrue
\mciteSetBstMidEndSepPunct{\mcitedefaultmidpunct}
{\mcitedefaultendpunct}{\mcitedefaultseppunct}\relax
\EndOfBibitem
\bibitem[Pham \latin{et~al.}(2013)Pham, Nguyen, Rocca, and Galli]{Pham2013}
Pham,~T.~A.; Nguyen,~H.-V.; Rocca,~D.; Galli,~G. GW calculations using the
  spectral decomposition of the dielectric matrix: Verification, validation,
  and comparison of methods. \emph{Phys. Rev. B} \textbf{2013}, \emph{87},
  155148\relax
\mciteBstWouldAddEndPuncttrue
\mciteSetBstMidEndSepPunct{\mcitedefaultmidpunct}
{\mcitedefaultendpunct}{\mcitedefaultseppunct}\relax
\EndOfBibitem
\bibitem[Govoni and Galli(2015)Govoni, and Galli]{Govoni2015}
Govoni,~M.; Galli,~G. Large scale GW calculations. \emph{Journal of chemical
  theory and computation} \textbf{2015}, \emph{11}, 2680--2696\relax
\mciteBstWouldAddEndPuncttrue
\mciteSetBstMidEndSepPunct{\mcitedefaultmidpunct}
{\mcitedefaultendpunct}{\mcitedefaultseppunct}\relax
\EndOfBibitem
\bibitem[Kaplan \latin{et~al.}(2016)Kaplan, Harding, Seiler, Weigend, Evers,
  and van Setten]{Kaplan2016}
Kaplan,~F.; Harding,~M.~E.; Seiler,~C.; Weigend,~F.; Evers,~F.; van
  Setten,~M.~J. Quasi-Particle Self-Consistent GW for Molecules. \emph{J. Chem.
  Theory Comput.} \textbf{2016}, \relax
\mciteBstWouldAddEndPunctfalse
\mciteSetBstMidEndSepPunct{\mcitedefaultmidpunct}
{}{\mcitedefaultseppunct}\relax
\EndOfBibitem
\bibitem[Heyd \latin{et~al.}(2003)Heyd, Scuseria, and Ernzerhof]{Heyd2003}
Heyd,~J.; Scuseria,~G.~E.; Ernzerhof,~M. Hybrid functionals based on a screened
  Coulomb potential. \emph{J. Chem. Phys.} \textbf{2003}, \emph{118},
  8207--8215\relax
\mciteBstWouldAddEndPuncttrue
\mciteSetBstMidEndSepPunct{\mcitedefaultmidpunct}
{\mcitedefaultendpunct}{\mcitedefaultseppunct}\relax
\EndOfBibitem
\bibitem[Baer \latin{et~al.}(2010)Baer, Livshits, and Salzner]{Baer2010a}
Baer,~R.; Livshits,~E.; Salzner,~U. Tuned Range-separated hybrids in density
  functional theory. \emph{Annu. Rev. Phys. Chem.} \textbf{2010}, \emph{61},
  85--109\relax
\mciteBstWouldAddEndPuncttrue
\mciteSetBstMidEndSepPunct{\mcitedefaultmidpunct}
{\mcitedefaultendpunct}{\mcitedefaultseppunct}\relax
\EndOfBibitem
\bibitem[Kronik \latin{et~al.}(2012)Kronik, Stein, Refaely-Abramson, and
  Baer]{Kronik2012}
Kronik,~L.; Stein,~T.; Refaely-Abramson,~S.; Baer,~R. Excitation Gaps of
  Finite-Sized Systems from Optimally Tuned Range-Separated Hybrid Functionals.
  \emph{J. Chem. Theory Comput.} \textbf{2012}, \emph{8}, 1515--1531\relax
\mciteBstWouldAddEndPuncttrue
\mciteSetBstMidEndSepPunct{\mcitedefaultmidpunct}
{\mcitedefaultendpunct}{\mcitedefaultseppunct}\relax
\EndOfBibitem
\bibitem[Neuhauser \latin{et~al.}(2015)Neuhauser, Rabani, Cytter, and
  Baer]{Neuhauser2015}
Neuhauser,~D.; Rabani,~E.; Cytter,~Y.; Baer,~R. Stochastic Optimally Tuned
  Range-Separated Hybrid Density Functional Theory. \emph{J. Phys. Chem. A}
  \textbf{2015}, \emph{120}, 3071--3078\relax
\mciteBstWouldAddEndPuncttrue
\mciteSetBstMidEndSepPunct{\mcitedefaultmidpunct}
{\mcitedefaultendpunct}{\mcitedefaultseppunct}\relax
\EndOfBibitem
\bibitem[Hedin(1965)]{Hedin1965}
Hedin,~L. New Method for Calculating the One-Particle Green's Function with
  Application to the Electron-Gas Problem. \emph{Phys. Rev.} \textbf{1965},
  \emph{139}, A796--A823\relax
\mciteBstWouldAddEndPuncttrue
\mciteSetBstMidEndSepPunct{\mcitedefaultmidpunct}
{\mcitedefaultendpunct}{\mcitedefaultseppunct}\relax
\EndOfBibitem
\bibitem[Hybertsen and Louie(1986)Hybertsen, and Louie]{Hybertsen1986}
Hybertsen,~M.~S.; Louie,~S.~G. Electron correlation in semiconductors and
  insulators: Band gaps and quasiparticle energies. \emph{Phys. Rev. B}
  \textbf{1986}, \emph{34}, 5390--5413\relax
\mciteBstWouldAddEndPuncttrue
\mciteSetBstMidEndSepPunct{\mcitedefaultmidpunct}
{\mcitedefaultendpunct}{\mcitedefaultseppunct}\relax
\EndOfBibitem
\bibitem[Aryasetiawan and Gunnarsson(1998)Aryasetiawan, and
  Gunnarsson]{Aryasetiawan1998}
Aryasetiawan,~F.; Gunnarsson,~O. The GW method. \emph{Rep. Prog. Phys.}
  \textbf{1998}, \emph{61}, 237--312\relax
\mciteBstWouldAddEndPuncttrue
\mciteSetBstMidEndSepPunct{\mcitedefaultmidpunct}
{\mcitedefaultendpunct}{\mcitedefaultseppunct}\relax
\EndOfBibitem
\bibitem[Onida \latin{et~al.}(2002)Onida, Reining, and Rubio]{Onida2002}
Onida,~G.; Reining,~L.; Rubio,~A. Electronic excitations: density-functional
  versus many-body Green's-function approaches. \emph{Rev. Mod. Phys.}
  \textbf{2002}, \emph{74}, 601--659\relax
\mciteBstWouldAddEndPuncttrue
\mciteSetBstMidEndSepPunct{\mcitedefaultmidpunct}
{\mcitedefaultendpunct}{\mcitedefaultseppunct}\relax
\EndOfBibitem
\bibitem[Friedrich and Schindlmayr(2006)Friedrich, and
  Schindlmayr]{Friedrich2006}
Friedrich,~C.; Schindlmayr,~A. Many-body perturbation theory: the GW
  approximation. \emph{NIC Series} \textbf{2006}, \emph{31}, 335\relax
\mciteBstWouldAddEndPuncttrue
\mciteSetBstMidEndSepPunct{\mcitedefaultmidpunct}
{\mcitedefaultendpunct}{\mcitedefaultseppunct}\relax
\EndOfBibitem
\bibitem[Rohlfing and Louie(2000)Rohlfing, and Louie]{Rohlfing2000}
Rohlfing,~M.; Louie,~S.~G. Electron-hole excitations and optical spectra from
  first principles. \emph{Phys. Rev. B} \textbf{2000}, \emph{62},
  4927--4944\relax
\mciteBstWouldAddEndPuncttrue
\mciteSetBstMidEndSepPunct{\mcitedefaultmidpunct}
{\mcitedefaultendpunct}{\mcitedefaultseppunct}\relax
\EndOfBibitem
\bibitem[Benedict \latin{et~al.}(2003)Benedict, Puzder, Williamson, Grossman,
  Galli, Klepeis, Raty, and Pankratov]{Benedict2003}
Benedict,~L.~X.; Puzder,~A.; Williamson,~A.~J.; Grossman,~J.~C.; Galli,~G.;
  Klepeis,~J.~E.; Raty,~J.-Y.; Pankratov,~O. Calculation of optical absorption
  spectra of hydrogenated Si clusters: Bethe-Salpeter equation versus
  time-dependent local-density approximation. \emph{Phys. Rev. B}
  \textbf{2003}, \emph{68}, 085310\relax
\mciteBstWouldAddEndPuncttrue
\mciteSetBstMidEndSepPunct{\mcitedefaultmidpunct}
{\mcitedefaultendpunct}{\mcitedefaultseppunct}\relax
\EndOfBibitem
\bibitem[Tiago and Chelikowsky(2006)Tiago, and Chelikowsky]{Tiago2006}
Tiago,~M.~L.; Chelikowsky,~J.~R. Optical excitations in organic molecules,
  clusters, and defects studied by first-principles Green function methods.
  \emph{Phys. Rev. B} \textbf{2006}, \emph{73}, 205334\relax
\mciteBstWouldAddEndPuncttrue
\mciteSetBstMidEndSepPunct{\mcitedefaultmidpunct}
{\mcitedefaultendpunct}{\mcitedefaultseppunct}\relax
\EndOfBibitem
\bibitem[Neuhauser \latin{et~al.}(2014)Neuhauser, Gao, Arntsen, Karshenas,
  Rabani, and Baer]{Neuhauser2014}
Neuhauser,~D.; Gao,~Y.; Arntsen,~C.; Karshenas,~C.; Rabani,~E.; Baer,~R.
  Breaking the Theoretical Scaling Limit for Predicting Quasiparticle Energies:
  The Stochastic G W Approach. \emph{Phys. Rev. Lett.} \textbf{2014},
  \emph{113}, 076402\relax
\mciteBstWouldAddEndPuncttrue
\mciteSetBstMidEndSepPunct{\mcitedefaultmidpunct}
{\mcitedefaultendpunct}{\mcitedefaultseppunct}\relax
\EndOfBibitem
\bibitem[Rabani \latin{et~al.}(2015)Rabani, Baer, and Neuhauser]{Rabani2015}
Rabani,~E.; Baer,~R.; Neuhauser,~D. Time-dependent stochastic Bethe-Salpeter
  approach. \emph{Phys. Rev. B} \textbf{2015}, \emph{91}, 235302\relax
\mciteBstWouldAddEndPuncttrue
\mciteSetBstMidEndSepPunct{\mcitedefaultmidpunct}
{\mcitedefaultendpunct}{\mcitedefaultseppunct}\relax
\EndOfBibitem
\bibitem[van Setten \latin{et~al.}(2015)van Setten, Caruso, Sharifzadeh, Ren,
  Scheffler, Liu, Lischner, Lin, Deslippe, Louie, Yang, Weigend, Neaton, Evers,
  and Rinke]{vanSetten2015}
van Setten,~M.~J.; Caruso,~F.; Sharifzadeh,~S.; Ren,~X.; Scheffler,~M.;
  Liu,~F.; Lischner,~J.; Lin,~L.; Deslippe,~J.~R.; Louie,~S.~G.; Yang,~C.;
  Weigend,~F.; Neaton,~J.~B.; Evers,~F.; Rinke,~P. GW 100: Benchmarking G 0 W 0
  for Molecular Systems. \emph{J. Chem. Theory Comput.} \textbf{2015},
  \emph{11}, 5665--5687\relax
\mciteBstWouldAddEndPuncttrue
\mciteSetBstMidEndSepPunct{\mcitedefaultmidpunct}
{\mcitedefaultendpunct}{\mcitedefaultseppunct}\relax
\EndOfBibitem
\bibitem[Furche \latin{et~al.}(2014)Furche, Ahlrichs, H{\"a}ttig, Klopper,
  Sierka, and Weigend]{furche2014turbomole}
Furche,~F.; Ahlrichs,~R.; H{\"a}ttig,~C.; Klopper,~W.; Sierka,~M.; Weigend,~F.
  Turbomole. \emph{Wiley Interdisciplinary Reviews: Computational Molecular
  Science} \textbf{2014}, \emph{4}, 91--100\relax
\mciteBstWouldAddEndPuncttrue
\mciteSetBstMidEndSepPunct{\mcitedefaultmidpunct}
{\mcitedefaultendpunct}{\mcitedefaultseppunct}\relax
\EndOfBibitem
\bibitem[Blum \latin{et~al.}(2009)Blum, Gehrke, Hanke, Havu, Havu, Ren, Reuter,
  and Scheffler]{Blum2009}
Blum,~V.; Gehrke,~R.; Hanke,~F.; Havu,~P.; Havu,~V.; Ren,~X.; Reuter,~K.;
  Scheffler,~M. Ab initio molecular simulations with numeric atom-centered
  orbitals. \emph{Computer Physics Communications} \textbf{2009}, \emph{180},
  2175--2196\relax
\mciteBstWouldAddEndPuncttrue
\mciteSetBstMidEndSepPunct{\mcitedefaultmidpunct}
{\mcitedefaultendpunct}{\mcitedefaultseppunct}\relax
\EndOfBibitem
\bibitem[Ren \latin{et~al.}(2012)Ren, Rinke, Blum, Wieferink, Tkatchenko,
  Sanfilippo, Reuter, and Scheffler]{Ren2012a}
Ren,~X.; Rinke,~P.; Blum,~V.; Wieferink,~J.; Tkatchenko,~A.; Sanfilippo,~A.;
  Reuter,~K.; Scheffler,~M. Resolution-of-identity approach to Hartree--Fock,
  hybrid density functionals, RPA, MP2 and GW with numeric atom-centered
  orbital basis functions. \emph{New Journal of Physics} \textbf{2012},
  \emph{14}, 053020\relax
\mciteBstWouldAddEndPuncttrue
\mciteSetBstMidEndSepPunct{\mcitedefaultmidpunct}
{\mcitedefaultendpunct}{\mcitedefaultseppunct}\relax
\EndOfBibitem
\bibitem[Nguyen \latin{et~al.}(2012)Nguyen, Pham, Rocca, and Galli]{Nguyen2012}
Nguyen,~H.-V.; Pham,~T.~A.; Rocca,~D.; Galli,~G. Improving accuracy and
  efficiency of calculations of photoemission spectra within the many-body
  perturbation theory. \emph{Phys. Rev. B} \textbf{2012}, \emph{85},
  081101\relax
\mciteBstWouldAddEndPuncttrue
\mciteSetBstMidEndSepPunct{\mcitedefaultmidpunct}
{\mcitedefaultendpunct}{\mcitedefaultseppunct}\relax
\EndOfBibitem
\bibitem[Baer \latin{et~al.}(2013)Baer, Neuhauser, and Rabani]{Baer2013}
Baer,~R.; Neuhauser,~D.; Rabani,~E. Self-Averaging Stochastic Kohn-Sham
  Density-Functional Theory. \emph{Phys. Rev. Lett.} \textbf{2013}, \emph{111},
  106402\relax
\mciteBstWouldAddEndPuncttrue
\mciteSetBstMidEndSepPunct{\mcitedefaultmidpunct}
{\mcitedefaultendpunct}{\mcitedefaultseppunct}\relax
\EndOfBibitem
\bibitem[Neuhauser \latin{et~al.}(2014)Neuhauser, Baer, and
  Rabani]{Neuhauser2014a}
Neuhauser,~D.; Baer,~R.; Rabani,~E. Communication: Embedded fragment stochastic
  density functional theory. \emph{J. Chem. Phys.} \textbf{2014}, \emph{141},
  041102\relax
\mciteBstWouldAddEndPuncttrue
\mciteSetBstMidEndSepPunct{\mcitedefaultmidpunct}
{\mcitedefaultendpunct}{\mcitedefaultseppunct}\relax
\EndOfBibitem
\bibitem[Hutchinson(1990)]{Hutchinson1990}
Hutchinson,~M.~F. A stochastic estimator of the trace of the influence matrix
  for Laplacian smoothing splines. \emph{Communications in
  Statistics-Simulation and Computation} \textbf{1990}, \emph{19},
  433--450\relax
\mciteBstWouldAddEndPuncttrue
\mciteSetBstMidEndSepPunct{\mcitedefaultmidpunct}
{\mcitedefaultendpunct}{\mcitedefaultseppunct}\relax
\EndOfBibitem
\bibitem[Baer and Head-Gordon(1997)Baer, and Head-Gordon]{Baer1997a}
Baer,~R.; Head-Gordon,~M. {Chebyshev} expansion methods for electronic
  structure calculations on large molecular systems. \emph{J. Chem. Phys.}
  \textbf{1997}, \emph{107}, 10003--10013\relax
\mciteBstWouldAddEndPuncttrue
\mciteSetBstMidEndSepPunct{\mcitedefaultmidpunct}
{\mcitedefaultendpunct}{\mcitedefaultseppunct}\relax
\EndOfBibitem
\bibitem[Baer and Head-Gordon(1997)Baer, and Head-Gordon]{Baer1997b}
Baer,~R.; Head-Gordon,~M. Sparsity of the density matrix in Kohn-Sham density
  functional theory and an assessment of linear system-size scaling methods.
  \emph{Phys. Rev. Lett.} \textbf{1997}, \emph{79}, 3962--3965\relax
\mciteBstWouldAddEndPuncttrue
\mciteSetBstMidEndSepPunct{\mcitedefaultmidpunct}
{\mcitedefaultendpunct}{\mcitedefaultseppunct}\relax
\EndOfBibitem
\bibitem[Baer and Neuhauser(2004)Baer, and Neuhauser]{Baer2004b}
Baer,~R.; Neuhauser,~D. Real-time linear response for time-dependent
  density-functional theory. \emph{J. Chem. Phys.} \textbf{2004}, \emph{121},
  9803--9807\relax
\mciteBstWouldAddEndPuncttrue
\mciteSetBstMidEndSepPunct{\mcitedefaultmidpunct}
{\mcitedefaultendpunct}{\mcitedefaultseppunct}\relax
\EndOfBibitem
\bibitem[Fetter and Walecka(1971)Fetter, and Walecka]{Fetter1971}
Fetter,~A.~L.; Walecka,~J.~D. \emph{Quantum Thoery of Many Particle Systems};
  McGraw-Hill: New York, 1971; p 299\relax
\mciteBstWouldAddEndPuncttrue
\mciteSetBstMidEndSepPunct{\mcitedefaultmidpunct}
{\mcitedefaultendpunct}{\mcitedefaultseppunct}\relax
\EndOfBibitem
\bibitem[Gao \latin{et~al.}(2015)Gao, Neuhauser, Baer, and Rabani]{Gao2015}
Gao,~Y.; Neuhauser,~D.; Baer,~R.; Rabani,~E. Sublinear scaling for
  time-dependent stochastic density functional theory. \emph{J. Chem. Phys.}
  \textbf{2015}, \emph{142}, 034106\relax
\mciteBstWouldAddEndPuncttrue
\mciteSetBstMidEndSepPunct{\mcitedefaultmidpunct}
{\mcitedefaultendpunct}{\mcitedefaultseppunct}\relax
\EndOfBibitem
\bibitem[Troullier and Martins(1991)Troullier, and Martins]{Troullier1991}
Troullier,~N.; Martins,~J.~L. Efficient Pseudopotentials for Plane-Wave
  Calculations. \emph{Phys. Rev. B} \textbf{1991}, \emph{43}, 1993--2006\relax
\mciteBstWouldAddEndPuncttrue
\mciteSetBstMidEndSepPunct{\mcitedefaultmidpunct}
{\mcitedefaultendpunct}{\mcitedefaultseppunct}\relax
\EndOfBibitem
\bibitem[{NIST Computational Chemistry Comparison and Benchmark Database NIST
  Standard Reference Database Number 101 Release 18, October 2016, Editor:
  Russell D. Johnson III http://cccbdb.nist.gov/}()]{NISTDataBase}
{NIST Computational Chemistry Comparison and Benchmark Database NIST Standard
  Reference Database Number 101 Release 18, October 2016, Editor: Russell D.
  Johnson III http://cccbdb.nist.gov/},\relax
\mciteBstWouldAddEndPuncttrue
\mciteSetBstMidEndSepPunct{\mcitedefaultmidpunct}
{\mcitedefaultendpunct}{\mcitedefaultseppunct}\relax
\EndOfBibitem
\bibitem[Vl{\v{c}}ek \latin{et~al.}(2017)Vl{\v{c}}ek, Baer, Rabani, and
  Neuhauser]{vlcek2017self}
Vl{\v{c}}ek,~V.; Baer,~R.; Rabani,~E.; Neuhauser,~D. Self-consistent band-gap
  renormalization GW. \emph{arXiv preprint arXiv:1701.02023} \textbf{2017},
  \relax
\mciteBstWouldAddEndPunctfalse
\mciteSetBstMidEndSepPunct{\mcitedefaultmidpunct}
{}{\mcitedefaultseppunct}\relax
\EndOfBibitem
\end{mcitethebibliography}
\end{document}